\documentclass{nature}
\usepackage{graphicx}
\usepackage{caption}
\title{Clouds in the atmosphere of the super-Earth exoplanet GJ\,1214b}

\author{Laura Kreidberg$^{1}$, Jacob L. Bean$^{1}$, Jean-Michel D\'{e}sert$^{2,3}$, Bj\"{o}rn Benneke$^{4}$,
Drake Deming$^5$, Kevin B. Stevenson$^1$, Sara Seager$^4$, Zachory Berta-Thompson$^{6,7}$, Andreas Seifahrt$^1$,
\& Derek Homeier$^8$}

\begin{document}

\maketitle

\begin{affiliations}
 \item Department of Astronomy and Astrophysics, University of Chicago, Chicago, IL 60637
 \item CASA, Department of Astrophysical \& Planetary Sciences, University of Colorado, Boulder, CO 80309
 \item Department of Astronomy, California Institute of Technology, Pasadena, CA 91101 
 \item Department of Physics, Massachussetts Insitute of Technology, Cambridge, MA 02139
 \item Department of Astronomy, University of Maryland, College Park, MD 20742
 \item Department of Astronomy, Harvard University, Cambridge, MA 02138
 \item MIT Kavli Institute for Astrophysics and Space Research, MIT, Cambridge, MA 02139
 \item Centre de Recherche Astrophysique de Lyon, ENS Lyon, Lyon, France
\end{affiliations}

\begin{abstract}
Recent surveys have revealed that planets intermediate in size between
Earth and Neptune (``super-Earths'') are among the most common planets
in the Galaxy\cite{cassan12, fressin13, petigura13}. 
Atmospheric studies are the next step toward developing
a comprehensive understanding of this new class of object\cite{adams08, kempton09,rogers10b}.
Much effort has been focused on using transmission spectroscopy to characterize the atmosphere of the 
super-Earth archetype GJ\,1214b\cite{bean10, desert11,bean11,
berta12,fraine13,kempton10,nettelmann11,kempton12,howe12,morley13,benneke13},
but previous observations did not have sufficient precision to distinguish between two interpretations for the atmosphere.
The planet's atmosphere could be dominated by relatively heavy molecules, such as water (e.g., a 100\% water vapor composition), or
it could contain high-altitude clouds that obscure its lower layers. Here we report a measurement of the
transmission spectrum of GJ\,1214b at near-infrared wavelengths that definitively resolves this ambiguity.
These data, obtained with the Hubble Space Telescope,
are sufficiently precise to detect absorption features from a high mean 
molecular mass atmosphere.  The observed spectrum, however, is featureless. We
rule out cloud-free atmospheric models with water-, methane-,
carbon monoxide-, nitrogen\mbox{-,} or carbon dioxide-dominated compositions at greater than
5\,$\sigma$ confidence.  The planet's atmosphere must contain clouds 
to be consistent with the data.
\end{abstract}

We observed 15 transits of the planet GJ\,1214b with the Wide Field Camera 3  (WFC3) instrument on the Hubble Space Telescope (HST) between UT 27 September 2012 and 22 August 2013.  Each transit observation consisted of four orbits of the telescope, with 45-minute gaps in phase coverage between target visibility periods due to Earth occultation.  We obtained time-series spectroscopy from 1.1 to 1.7\,$\mu$m during each observation.  The data were taken in spatial scan mode, which slews the telescope during the exposure and moves the spectrum perpendicular to the dispersion direction on the detector.  This mode reduces the instrumental overhead time by a factor of five compared to staring mode observations.  We achieved an integration efficiency of 60 -- 70\%.  We extracted the spectra and divided each exposure into five-pixel-wide bins, obtaining spectro-photometric time series in 22 channels (resolution $R\equiv \lambda/\Delta\lambda \sim 70$).  The typical signal-to-noise per 88.4\,s exposure per channel was 1,400.  We also created a ``white'' light curve summed over the entire wavelength range.  
Our analysis incorporates data from 12 of the 15 transits observed, because one observation was compromised due to a telescope guiding error and two showed evidence of a starspot crossing.  

The raw transit light curves for GJ 1214b exhibit ramp-like systematics comparable to those seen in previous WFC3 data\cite{berta12,deming13,swain13}.  The ramp in the first orbit of each visit consistently has the largest amplitude and a different shape from ramps in the subsequent orbits.
Following standard  procedure for HST transit light curves, we did not include data from the first orbit in our analysis, leaving 654 exposures.  We corrected for systematics in the remaining three orbits using two techniques that have been successfully applied in prior analyses\cite{berta12,deming13,stevenson13}.  The first approach models the systematics as an analytic function of time.  The function includes an exponential ramp term fit to each orbit, a visit-long slope, and a normalization factor.  The second approach assumes the morphology of the systematics is independent of wavelength, and models each channel with a scalar multiple of the time series of systematics from the white light curve fit.  We obtained consistent results from both methods (see Extended Data Table 1), and report here results from the second.  See the Supplementary Information and Extended Data Figs. 1 -- 6 for more detail on the observations, data reduction, and systematics correction.

We fit the light curves in each spectroscopic channel with a transit model\cite{mandel02} to measure the transit depth as a function of wavelength; this constitutes the transmission spectrum.  See Figure 1 for the fitted transit light curves.  We used the second systematics correction technique described above and fit a unique planet-to-star radius ratio $R_p/R_s$ and normalization $C$ to each channel and each visit, and a unique linear limb darkening parameter $u$ to each channel.  We assumed a circular orbit\cite{anglada13} and fixed the inclination $i = 89.1^\circ$, the ratio of the semi-major axis to the stellar radius $a/R_s = 15.23$, the orbital period $P =1.58040464894$\,days, and the time of central transit $T_c = 2454966.52488\, \mathrm{BJD_{TDB}}$. These are the best fit values to the white light curve.  

The measured transit depths in each channel are consistent over all transit epochs (see Extended Data Fig. 5), and we report the weighted average depth per channel.  The resulting transmission spectrum is shown in Figure 2.  Our results are not significantly affected by stellar activity, as we discuss further in the Supplemental Information.  Careful treatment of the limb darkening is critical to the results,
but our limb darkening measurements are not degenerate with the transit depth (see Extended Data Fig. 4) and agree with the predictions from theoretical models (see Extended Data Fig. 6).  
Our conclusions are unchanged if we fix the limb darkening on theoretical values.  We find that a linear limb darkening law is sufficient to model the data.  For further description of the limb darkening treatment, see the Supplementary Information.

The transmission spectrum we report here has the precision necessary to
detect the spectral features of a high mean molecular mass atmosphere for the first time. 
However, the observed spectrum is featureless.  The data are best fit with a flat line, which has a 
reduced $\chi^2$ of 1.0.  
We compare several models to the data that represent limiting case scenarios in the range of
expected atmospheric compositions\cite{benneke13,fortney13}.  Depending on the formation history
and evolution of the planet, a high mean molecular mass atmosphere could be dominated by
water (H$_2$O), methane (CH$_4$), carbon monoxide (CO), carbon dioxide (CO$_2$), or
nitrogen (N$_2$).  Water is expected to be the dominant absorber in the wavelength
range of our observations, so a wide range of high mean molecular mass atmospheres with
trace amounts of water can be approximated by a pure H$_2$O model.  The data show no evidence
for water absorption.  A cloud-free pure H$_2$O composition is ruled out at 16.1\,$\sigma$ confidence.  
In the case of a dry atmosphere, features from other absorbers such as CH$_4$, CO, or
CO$_2$ could be visible in the transmission spectrum.  Cloud-free atmospheres composed of these absorbers are
also excluded by the data, at 31.1, 7.5, and 5.5\,$\sigma$ confidence, respectively.
Nitrogen has no spectral features in the observed wavelength range, but
our measurements are sensitive to a nitrogen-rich atmosphere with trace amounts of spectrally active
molecules.  For example, we can rule out a 99.9\% N$_2$, 0.1\% H$_2$O atmosphere at 5.6\,$\sigma$ confidence.
Of the scenarios considered here, a 100\% CO$_2$ atmosphere is the most challenging to detect because CO$_2$ has the highest molecular mass 
and a relatively small opacity in the observed wavelength range.  
Given that the data are precise enough to rule out even a CO$_2$ composition at high confidence,
the most likely explanation for the absence of spectral features is a gray opacity source, suggesting
that clouds are present in the atmosphere.  Clouds can block transmission of stellar flux through
the atmosphere, which truncates spectral features arising from below the cloud altitude\cite{fortney05}.

To illustrate the properties of potential clouds, we perform a Bayesian analysis 
on the transmission spectrum with a code designed for spectral retrieval of super-Earth atmospheric compositions\cite{benneke12}. 
 We assume a two-component model atmosphere of water and a solar mix of hydrogen/helium gas, motivated by the fact
that water is the most abundant icy volatile for solar abundance ratios.
Clouds are modeled as a gray, optically thick opacity source below a given altitude.
See Figure 3 for the retrieval results.  For this model, the data constrain the cloud top pressure to less than
 $10^{-2}$\,mbar for a mixing ratio with mean molecular mass equal to solar and less than  $10^{-1}$\,mbar for a water-dominated 
composition 
(both at 3\,$\sigma$ confidence). At the temperatures and pressures expected in the atmosphere of GJ\,1214b,
equilibrium condensates of ZnS and KCl can form in the observable part of the atmosphere. 
While these species  
could provide the necessary opacity, they are predicted to form at much higher pressures
(deeper than 10\,mbar for a 50x solar metallicity model)\cite{morley13}, requiring that 
clouds be lofted high from their base altitude to explain our measured spectrum.
Alternatively, photochemistry could produce a layer of hydrocarbons in the upper atmosphere,
analogous to the haze on Saturn's moon Titan\cite{kempton12, morley13}.

The result presented here demonstrates the capability of current
facilities to measure very precise spectra of exoplanets by combining many
transit observations.  This observational strategy has the potential to yield the
atmospheric characterization of an Earth-size planet orbiting in the habitable zone of
a small, nearby star. 
Transmission spectrum features probing five scale heights of a nitrogen-rich atmosphere
on such a planet would have an amplitude of 30 ppm,
which is comparable to the photon-limited measurement precision we obtained with the Hubble Space Telescope.  
However, our findings for the super-Earth archetype GJ\,1214b, as well as emerging results for
hot, giant exoplanets\cite{pont08,deming13}, suggest that clouds may exist 
across a wide range of planetary atmosphere compositions,
temperatures, and pressures.  Clouds generally do not have constant opacity at all wavelengths,
so further progress in this area can be made by 
obtaining high-precision data with broad spectral coverage. 
Another avenue forward is to focus on measuring exoplanet emission and reflection spectra during
secondary eclipse, because the optical depth of clouds viewed at near-normal incidence is lower 
than that for the slant geometry observed during transit\cite{fortney05}.
Fortunately, the next generation of large
ground-based telescopes and the James Webb Space Telescope will have the
capabilities to make these kinds of measurements,
bringing us within reach of characterizing potentially habitable worlds beyond 
our Solar System.

\bibliographystyle{naturemag}
\bibliography{ms.bib}

\begin{thebibliography}{10}
\expandafter\ifx\csname url\endcsname\relax
  \def\url#1{\texttt{#1}}\fi
\expandafter\ifx\csname urlprefix\endcsname\relax\def\urlprefix{URL }\fi
\providecommand{\bibinfo}[2]{#2}
\providecommand{\eprint}[2][]{\url{#2}}

\bibitem{cassan12}
\bibinfo{author}{{Cassan}, A.} \emph{et~al.}
\newblock \bibinfo{title}{{One or more bound planets per Milky Way star from
  microlensing observations}}.
\newblock \emph{\bibinfo{journal}{Nature}} \textbf{\bibinfo{volume}{481}},
  \bibinfo{pages}{167--169} (\bibinfo{year}{2012}).

\bibitem{fressin13}
\bibinfo{author}{{Fressin}, F.} \emph{et~al.}
\newblock \bibinfo{title}{{The False Positive Rate of Kepler and the Occurrence
  of Planets}}.
\newblock \emph{\bibinfo{journal}{Astrophys.~J.~}}
  \textbf{\bibinfo{volume}{766}}, \bibinfo{pages}{81} (\bibinfo{year}{2013}).

\bibitem{petigura13}
\bibinfo{author}{{Petigura}, E.~A.}, \bibinfo{author}{{Marcy}, G.~W.} \&
  \bibinfo{author}{{Howard}, A.~W.}
\newblock \bibinfo{title}{{A Plateau in the Planet Population below Twice the
  Size of Earth}}.
\newblock \emph{\bibinfo{journal}{Astrophys.~J.~}}
  \textbf{\bibinfo{volume}{770}}, \bibinfo{pages}{69} (\bibinfo{year}{2013}).

\bibitem{adams08}
\bibinfo{author}{{Adams}, E.~R.}, \bibinfo{author}{{Seager}, S.} \&
  \bibinfo{author}{{Elkins-Tanton}, L.}
\newblock \bibinfo{title}{{Ocean Planet or Thick Atmosphere: On the Mass-Radius
  Relationship for Solid Exoplanets with Massive Atmospheres}}.
\newblock \emph{\bibinfo{journal}{Astrophys.~J.}}
  \textbf{\bibinfo{volume}{673}}, \bibinfo{pages}{1160--1164}
  (\bibinfo{year}{2008}).

\bibitem{kempton09}
\bibinfo{author}{{Miller-Ricci}, E.}, \bibinfo{author}{{Seager}, S.} \&
  \bibinfo{author}{{Sasselov}, D.}
\newblock \bibinfo{title}{{The Atmospheric Signatures of Super-Earths: How to
  Distinguish Between Hydrogen-Rich and Hydrogen-Poor Atmospheres}}.
\newblock \emph{\bibinfo{journal}{Astrophys.~J.}}
  \textbf{\bibinfo{volume}{690}}, \bibinfo{pages}{1056--1067}
  (\bibinfo{year}{2009}).

\bibitem{rogers10b}
\bibinfo{author}{{Rogers}, L.~A.} \& \bibinfo{author}{{Seager}, S.}
\newblock \bibinfo{title}{{Three Possible Origins for the Gas Layer on GJ
  1214b}}.
\newblock \emph{\bibinfo{journal}{Astrophys.~J.~}}
  \textbf{\bibinfo{volume}{716}}, \bibinfo{pages}{1208--1216}
  (\bibinfo{year}{2010}).

\bibitem{bean10}
\bibinfo{author}{{Bean}, J.~L.}, \bibinfo{author}{{Miller-Ricci Kempton}, E.}
  \& \bibinfo{author}{{Homeier}, D.}
\newblock \bibinfo{title}{{A ground-based transmission spectrum of the
  super-Earth exoplanet GJ 1214b}}.
\newblock \emph{\bibinfo{journal}{Nature}} \textbf{\bibinfo{volume}{468}},
  \bibinfo{pages}{669--672} (\bibinfo{year}{2010}).

\bibitem{desert11}
\bibinfo{author}{{D{\'e}sert}, J.-M.} \emph{et~al.}
\newblock \bibinfo{title}{{Observational Evidence for a Metal-rich Atmosphere
  on the Super-Earth GJ1214b}}.
\newblock \emph{\bibinfo{journal}{Astrophys.~J.~}}
  \textbf{\bibinfo{volume}{731}}, \bibinfo{pages}{L40} (\bibinfo{year}{2011}).

\bibitem{bean11}
\bibinfo{author}{{Bean}, J.~L.} \emph{et~al.}
\newblock \bibinfo{title}{{The Optical and Near-infrared Transmission Spectrum
  of the Super-Earth GJ 1214b: Further Evidence for a Metal-rich Atmosphere}}.
\newblock \emph{\bibinfo{journal}{Astrophys.~J.~}}
  \textbf{\bibinfo{volume}{743}}, \bibinfo{pages}{92} (\bibinfo{year}{2011}).

\bibitem{berta12}
\bibinfo{author}{{Berta}, Z.~K.} \emph{et~al.}
\newblock \bibinfo{title}{{The Flat Transmission Spectrum of the Super-Earth
  GJ1214b from Wide Field Camera 3 on the Hubble Space Telescope}}.
\newblock \emph{\bibinfo{journal}{Astrophys.~J.~}}
  \textbf{\bibinfo{volume}{747}}, \bibinfo{pages}{35} (\bibinfo{year}{2012}).

\bibitem{fraine13}
\bibinfo{author}{{Fraine}, J.~D.} \emph{et~al.}
\newblock \bibinfo{title}{{Spitzer Transits of the Super-Earth GJ1214b and
  Implications for its Atmosphere}}.
\newblock \emph{\bibinfo{journal}{Astrophys.~J.~}}
  \textbf{\bibinfo{volume}{765}}, \bibinfo{pages}{127} (\bibinfo{year}{2013}).

\bibitem{kempton10}
\bibinfo{author}{{Miller-Ricci}, E.} \& \bibinfo{author}{{Fortney}, J.~J.}
\newblock \bibinfo{title}{{The Nature of the Atmosphere of the Transiting
  Super-Earth GJ 1214b}}.
\newblock \emph{\bibinfo{journal}{Astrophys.~J.}}
  \textbf{\bibinfo{volume}{716}}, \bibinfo{pages}{L74--L79}
  (\bibinfo{year}{2010}).

\bibitem{nettelmann11}
\bibinfo{author}{{Nettelmann}, N.}, \bibinfo{author}{{Fortney}, J.~J.},
  \bibinfo{author}{{Kramm}, U.} \& \bibinfo{author}{{Redmer}, R.}
\newblock \bibinfo{title}{{Thermal Evolution and Structure Models of the
  Transiting Super-Earth GJ 1214b}}.
\newblock \emph{\bibinfo{journal}{Astrophys.~J.}}
  \textbf{\bibinfo{volume}{733}}, \bibinfo{pages}{2} (\bibinfo{year}{2011}).

\bibitem{kempton12}
\bibinfo{author}{{Miller-Ricci Kempton}, E.}, \bibinfo{author}{{Zahnle}, K.} \&
  \bibinfo{author}{{Fortney}, J.~J.}
\newblock \bibinfo{title}{{The Atmospheric Chemistry of GJ 1214b:
  Photochemistry and Clouds}}.
\newblock \emph{\bibinfo{journal}{Astrophys.~J.}}
  \textbf{\bibinfo{volume}{745}}, \bibinfo{pages}{3} (\bibinfo{year}{2012}).

\bibitem{howe12}
\bibinfo{author}{{Howe}, A.~R.} \& \bibinfo{author}{{Burrows}, A.~S.}
\newblock \bibinfo{title}{{Theoretical Transit Spectra for GJ 1214b and Other
  ''Super-Earths''}}.
\newblock \emph{\bibinfo{journal}{Astrophys.~J.}}
  \textbf{\bibinfo{volume}{756}}, \bibinfo{pages}{176} (\bibinfo{year}{2012}).

\bibitem{morley13}
\bibinfo{author}{{Morley}, C.~V.} \emph{et~al.}
\newblock \bibinfo{title}{{Quantitatively Assessing the Role of Clouds in the
  Transmission Spectrum of GJ 1214b}}.
\newblock \emph{\bibinfo{journal}{Astrophys.~J.}}
  \textbf{\bibinfo{volume}{775}}, \bibinfo{pages}{33} (\bibinfo{year}{2013}).

\bibitem{benneke13}
\bibinfo{author}{{Benneke}, B.} \& \bibinfo{author}{{Seager}, S.}
\newblock \bibinfo{title}{{How to Distinguish between Cloudy Mini-Neptunes and
  Water/Volatile-Dominated Super-Earths}}.
\newblock \emph{\bibinfo{journal}{ArXiv e-prints}}  (\bibinfo{year}{2013}).
\newblock \eprint{1306.6325}.

\bibitem{deming13}
\bibinfo{author}{{Deming}, D.} \emph{et~al.}
\newblock \bibinfo{title}{{Infrared Transmission Spectroscopy of the Exoplanets
  HD 209458b and XO-1b Using the Wide Field Camera-3 on the Hubble Space
  Telescope}}.
\newblock \emph{\bibinfo{journal}{Astrophys.~J.}}
  \textbf{\bibinfo{volume}{774}}, \bibinfo{pages}{95} (\bibinfo{year}{2013}).

\bibitem{swain13}
\bibinfo{author}{{Swain}, M.} \emph{et~al.}
\newblock \bibinfo{title}{{Probing the extreme planetary atmosphere of
  WASP-12b}}.
\newblock \emph{\bibinfo{journal}{Icarus}} \textbf{\bibinfo{volume}{225}},
  \bibinfo{pages}{432--445} (\bibinfo{year}{2013}).

\bibitem{stevenson13}
\bibinfo{author}{{Stevenson}, K.~B.} \emph{et~al.}
\newblock \bibinfo{title}{{Transmission Spectroscopy of the Hot-Jupiter
  WASP-12b from 0.7 to 5 microns}}.
\newblock \emph{\bibinfo{journal}{ArXiv e-prints}}  (\bibinfo{year}{2013}).
\newblock \eprint{1305.1670}.

\bibitem{mandel02}
\bibinfo{author}{{Mandel}, K.} \& \bibinfo{author}{{Agol}, E.}
\newblock \bibinfo{title}{{Analytic Light Curves for Planetary Transit
  Searches}}.
\newblock \emph{\bibinfo{journal}{Astrophys.~J.}}
  \textbf{\bibinfo{volume}{580}}, \bibinfo{pages}{L171--L175}
  (\bibinfo{year}{2002}).

\bibitem{anglada13}
\bibinfo{author}{{Anglada-Escud{\'e}}, G.}, \bibinfo{author}{{Rojas-Ayala},
  B.}, \bibinfo{author}{{Boss}, A.~P.}, \bibinfo{author}{{Weinberger}, A.~J.}
  \& \bibinfo{author}{{Lloyd}, J.~P.}
\newblock \bibinfo{title}{{GJ 1214 reviewed. Trigonometric parallax, stellar
  parameters, new orbital solution, and bulk properties for the super-Earth GJ
  1214b}}.
\newblock \emph{\bibinfo{journal}{Astron.~Astrophys.}}
  \textbf{\bibinfo{volume}{551}}, \bibinfo{pages}{A48} (\bibinfo{year}{2013}).

\bibitem{fortney13}
\bibinfo{author}{{Fortney}, J.~J.} \emph{et~al.}
\newblock \bibinfo{title}{{A Framework for Characterizing the Atmospheres of
  Low-mass Low-density Transiting Planets}}.
\newblock \emph{\bibinfo{journal}{Astrophys.~J.}}
  \textbf{\bibinfo{volume}{775}}, \bibinfo{pages}{80} (\bibinfo{year}{2013}).

\bibitem{fortney05}
\bibinfo{author}{{Fortney}, J.~J.}
\newblock \bibinfo{title}{{The effect of condensates on the characterization of
  transiting planet atmospheres with transmission spectroscopy}}.
\newblock \emph{\bibinfo{journal}{Mon.~Not.~R.~Astron.~Soc.}}
  \textbf{\bibinfo{volume}{364}}, \bibinfo{pages}{649--653}
  (\bibinfo{year}{2005}).

\bibitem{benneke12}
\bibinfo{author}{{Benneke}, B.} \& \bibinfo{author}{{Seager}, S.}
\newblock \bibinfo{title}{{Atmospheric Retrieval for Super-Earths: Uniquely
  Constraining the Atmospheric Composition with Transmission Spectroscopy}}.
\newblock \emph{\bibinfo{journal}{Astrophys.~J.}}
  \textbf{\bibinfo{volume}{753}}, \bibinfo{pages}{100} (\bibinfo{year}{2012}).

\bibitem{pont08}
\bibinfo{author}{{Pont}, F.}, \bibinfo{author}{{Knutson}, H.},
  \bibinfo{author}{{Gilliland}, R.~L.}, \bibinfo{author}{{Moutou}, C.} \&
  \bibinfo{author}{{Charbonneau}, D.}
\newblock \bibinfo{title}{{Detection of atmospheric haze on an extrasolar
  planet: the 0.55-1.05 {$\mu$}m transmission spectrum of HD 189733b with the
  Hubble Space Telescope}}.
\newblock \emph{\bibinfo{journal}{Mon.~Not.~R.~Astron.~Soc.}}
  \textbf{\bibinfo{volume}{385}}, \bibinfo{pages}{109--118}
  (\bibinfo{year}{2008}).

\end{thebibliography}

\begin{addendum}
\item [Supplementary Information] is available in the online version of the paper. 
\item [Acknowledgements] This work is based on observations made with the
NASA/ESA Hubble Space Telescope that were obtained at the Space Telescope
Science Institute, which is operated by the Association of Universities
for Research in Astronomy, Inc., under NASA contract NAS 5-26555. These
observations are associated with program GO-13021. Support for this work
was provided by NASA through a grant from the Space Telescope Science
Institute, the National Science Foundation through a Graduate Research
Fellowship (to L.K.), the Alfred P. Sloan Foundation through a Sloan
Research Fellowship (to J.L.B.), NASA through a Sagan Fellowship (to
J.-M.D.), and the European Research Council (for D.H. under 
the European Community's Seventh Framework Programme, FP7/2007-2013 Grant Agreement no. 247060).
\item[Author Contributions] L.K. led the data analysis, with contributions
from J.L.B., D.D., K.B.S., and A.S.; L.K., J.L.B, J.-M.D., and B.B. wrote the paper; J.L.B
and J.-M.D. conceived the project and wrote the telescope time proposal
with contributions from B.B., D.D., S.S., and Z.B.-T.; L.K., J.L.B., J.-M.D.,
D.D., and Z.B.-T. planned the observations; B.B. and S.S. developed and
performed the theoretical modeling; D.H. calculated theoretical stellar
limb darkening; J.L.B. led the overall direction of the project. All
authors discussed the results and commented on the manuscript. 
\item[Author Information] The data utilized in this work can be accessed at
the NASA Mikulski Archive for Space Telescopes (http://archive.stsci.edu).
Reprints and permissions information is available at
www.nature.com/reprints. The authors declare that they have no competing
financial interests. Correspondence and request for materials should be
addressed to L.K. (laura.kreidberg@uchicago.edu).
\end{addendum}

\begin{figure*}
\resizebox{16.0cm}{!}{\includegraphics{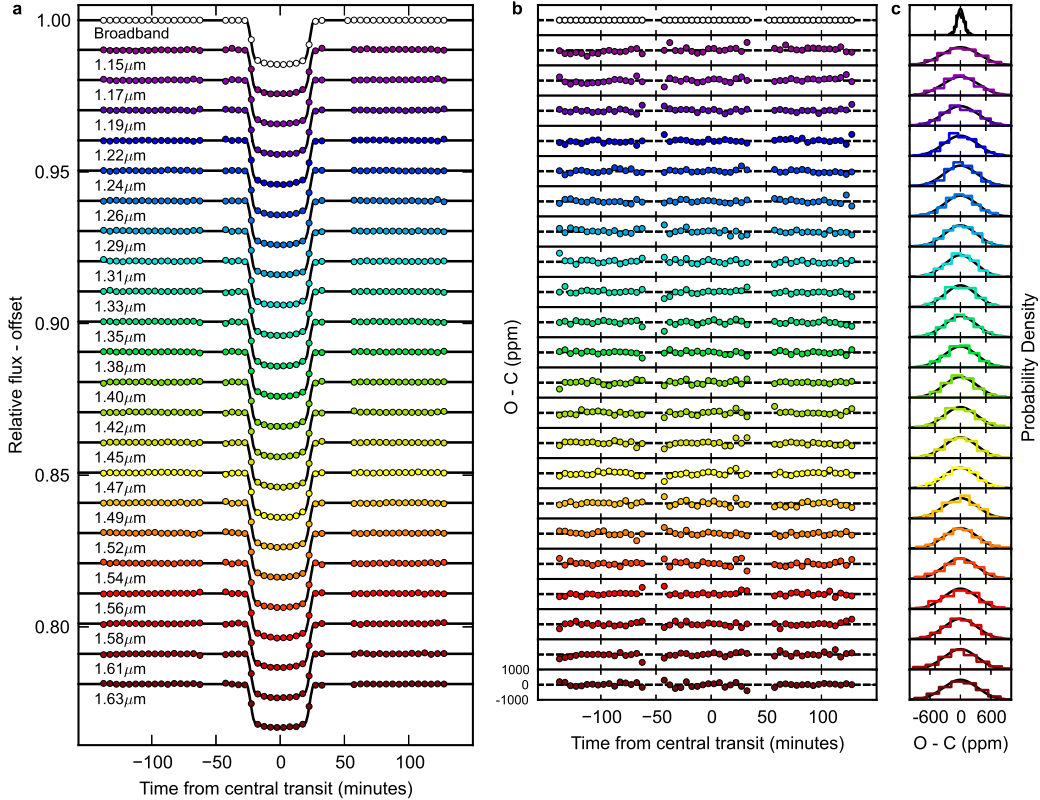}}
\caption{\textbf{Spectrophotometric data for transit observations of GJ\,1214b.}  \textbf{a}, Normalized and systematics-corrected data (points) with best-fit transit models (lines), offset for clarity.  The data consist of 12 transit observations and are binned in phase in 5-minute increments.  The spectroscopic light curve fit parameters are transit depth, a linear limb darkening coefficient, and a normalization term to correct for systematics. A unique transit depth is determined for each observation and the measured transit depths are consistent from epoch to epoch in all channels.  \textbf{b}, Binned residuals from the best-fit model light curves.  The residuals are within 14\% of the predicted photon-limited shot noise in all spectroscopic channels.  The median observed rms in the spectroscopic channels is 315 ppm, prior to binning.  \textbf{c}, Histograms of the unbinned residuals (colored lines) compared to the expected photon noise (black lines).  The residuals are Gaussian, satisfying a Shapiro-Wilk test for normality at the the $\alpha=0.1$ level in all but one channel (1.24\,$\mu$m).  The median reduced $\chi^2$ value for the spectroscopic light curve fits is 1.02.}
\end{figure*}

\begin{figure*}
\resizebox{14.0cm}{!}{\includegraphics{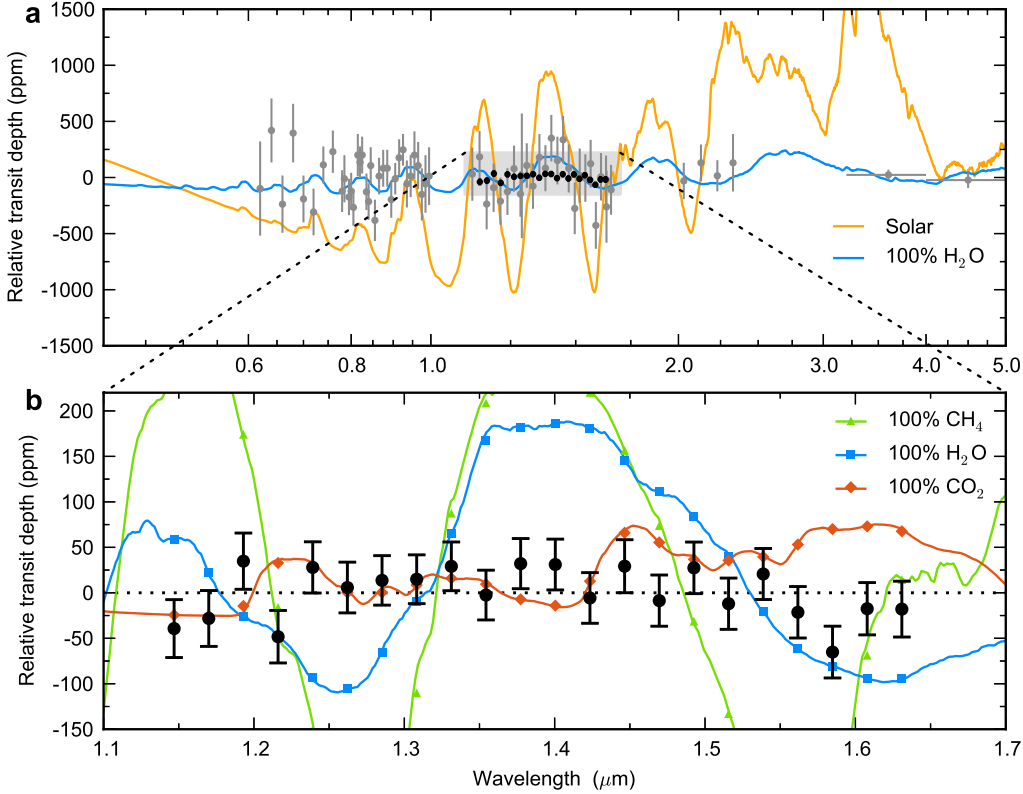}}
\caption{\textbf{The transmission spectrum of GJ\,1214b.}  \textbf{a}, Transmission spectrum measurements from our data (black points) and previous work (gray points)\cite{bean10, desert11,bean11,berta12,fraine13}, compared to theoretical models (lines).  The error bars correspond to 1\,$\sigma$ uncertainties.  Each data set is plotted relative to its mean.  Our measurements are consistent with past results for GJ\,1214 using WFC3\cite{berta12}.  Previous data rule out a cloud-free solar composition (orange line), but are consistent with a high-mean molecular weight atmosphere (e.g. 100\% water, blue line) or a hydrogen-rich atmosphere with high-altitude clouds.  \textbf{b}, Detail view of our measured transmission spectrum (black points) compared to high mean molecular weight models (lines).  The error bars are 1\,$\sigma$ uncertainties in the posterior distribution from a Markov chain Monte Carlo fit to the light curves (see the Supplemental Information for details of the fits).  The colored points correspond to the models binned at the resolution of the observations.  The data are consistent with a featureless spectrum ($\chi^2$ = 21.1 for 21 degrees of freedom), but inconsistent with cloud-free high-mean molecular weight scenarios.  Fits to pure water (blue line), methane (green line), carbon monoxide (not shown), and carbon dioxide (red line) models have $\chi^2 =$ 334.7, 1067.0, 110.0, and 75.4 with 21 degrees of freedom, and are ruled out at 16.1, 31.1, 7.5, and 5.5\,$\sigma$ confidence, respectively.}
\end{figure*}

\begin{figure*}
\resizebox{12.0cm}{!}{\includegraphics{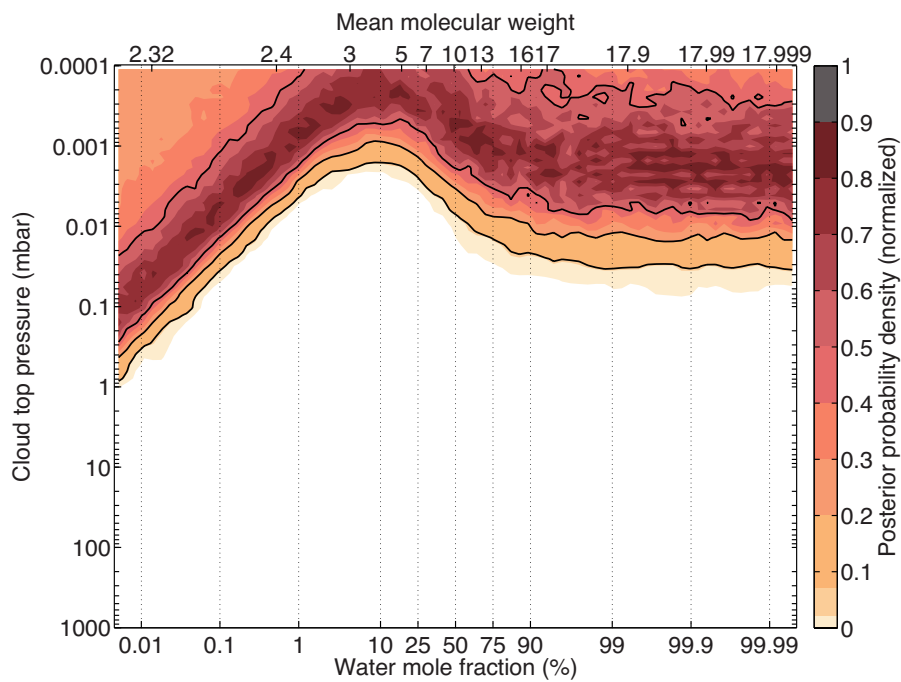}}
\caption{\textbf{Spectral retrieval results for a two-component (hydrogen/helium and water) model atmosphere for GJ\,1214b.} The colors indicate posterior probability density as a function of water mole fraction and cloud top pressure.  Black contours mark the 1, 2, and 3\,$\sigma$ Bayesian credible regions.  Clouds are modeled with a gray opacity, with transmission truncated below the cloud altitude.  The atmospheric modeling assumes a surface gravity of 8.48\,m/s$^2$ and an equilibrium temperature equal to 580\,K.}
\end{figure*}

\newpage

\begin{center}
{\LARGE{\bf Supplementary Information}}
\end{center}

The supplementary information describes the observations, data reduction, systematics correction, and light curve fitting for transit observations of the super-Earth GJ\,1214b. 

\vspace{-4mm}

\section*{Observations}\vspace{-4mm}
We observed 15 transits of the super-Earth exoplanet GJ\,1214b with the Wide Field Camera 3 (WFC3) instrument on the Hubble Space Telescope (HST) between UT 27 September 2012 and 20 August 2013.    Each transit observation (or visit) consisted of four 96-minute HST orbits of time series spectroscopy, with 45-minute gaps in data collection in each orbit due to Earth occultation.  We employed the G141 grism, which covers the wavelength range 1.1 to 1.7\,$\mu$m.  The spectra were binned  at resolution $R \equiv \lambda/\Delta\lambda \sim 70$.  To optimize the efficiency of the observations, we used spatial scan mode, which moves the spectrum perpendicular to the dispersion direction during the exposure.  Spatial scanning enables longer exposures for bright targets that would otherwise saturate, such as GJ\,1214.  We used a 0.12''/second scan rate for all exposures, which yielded peak per pixel counts near 23,000 electrons (30\% of saturation).  An example raw data frame is shown in Extended Data Figure 1.

\begin{figure*}[h!]
\centering
\includegraphics{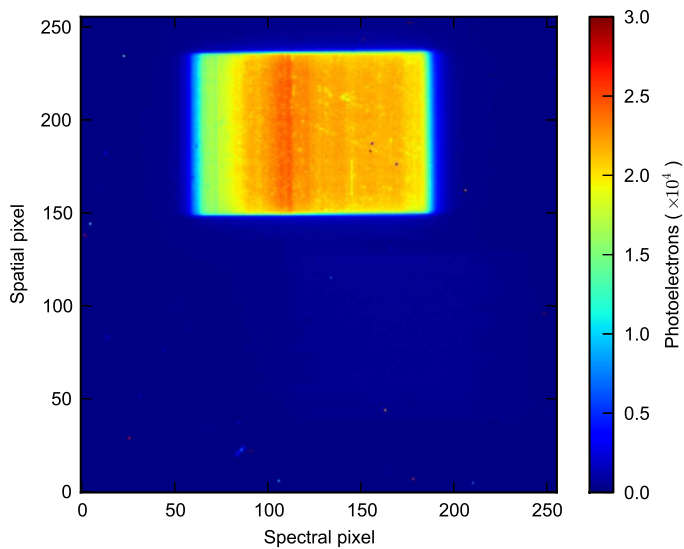}
\caption*{\textbf{Extended Data Figure 1:} An example spatially scanned raw data frame.  The exposure time was 88.4 s.}
\end{figure*}
 
The observations had the following design.  At the beginning of each orbit, we took a direct image with the F130N narrowband filter to establish a wavelength zero-point.  For the remainder of each orbit, we took spatially scanned exposures with the G141 grism.  Each observation used the $256 \times 256$ subarray.  During the first five transit observations, we took 88.4\,s exposures with the SPARS10, NSAMP=13 readout mode and scanned in the forward direction only.  Each exposure contains NSAMP non-destructive reads.  For transit observations 6 -- 15, we modified our approach to reduce overhead time: we increased the exposure time to 103.1\,s using the mode SPARS10, NSAMP=15, and scanned successively forward and backward.  These approaches yielded 67 and 75 spectra per visit with duty cycles of 58\% and  76\%, respectively.  One transit observation (UT 12 April 2013) was unsuccessful because the Fine Guidance Sensors failed to acquire the guide stars.  We do not use data from this observation in our analysis.  We also exclude data from the transit observations on UT 4 August 2013 and UT 12 August 2013, which showed evidence for a starspot crossing.  Our final analysis therefore used 12 transit observations.

\vspace{-4mm}

\section*{Data reduction}\vspace{-4mm}
Our data reduction process begins with the ``\texttt{ima}'' data product from the WFC3 calibration pipeline, \texttt{calwf3}.  These files are bias- and dark current-subtracted and flagged for bad pixels.  For spatially scanned data, each pixel is illuminated by the stellar spectrum for only a small fraction of the exposure; the remainder of time it collects background.  To aid in removing the background, we form subexposures of each image by subtracting consecutive non-destructive reads.  A subexposure thus contains photoelectrons gathered during the 7.4\,s between two reads.  We reduce each subexposure independently, as follows.   First we apply a wavelength-dependent flat field correction.  Next we mask bad pixels that have been flagged data quality DQ = 4, 32, or 512 by \texttt{calwf3}.  To estimate the background collected during the subexposure, we draw conservative masks around all stellar spectra, measure the background from the median of the unmasked pixels, and subtract it.  We compute a variance for the spectrum accounting for photon shot noise, detector read noise, and uncertainty in the background estimation.

We next correct for the wavelength dependence of the spectrum on detector position.  The grism dispersion varies along the spatial direction of the detector, so we calculate the dispersion solution for each row in the subexposure and interpolate the photoelectron counts in that row to the wavelength scale corresponding to the direct image position.  This interpolation also corrects bad pixels.  We then create a 40-pixel tall extraction box centered on the middle of the spatial scan and extract the spectrum with an optimal extraction routine.  Because each row has been interpolated to a common wavelength scale, the final spectrum is constructed by summing by column the spectra from all the subexposures.  The unit of time sampling in the light curve is thus a single exposure, which is the sum of 12 subexposures.  See Extended Data Figure 2 for an example extracted spectrum.  

\begin{figure*}[h!]
\includegraphics{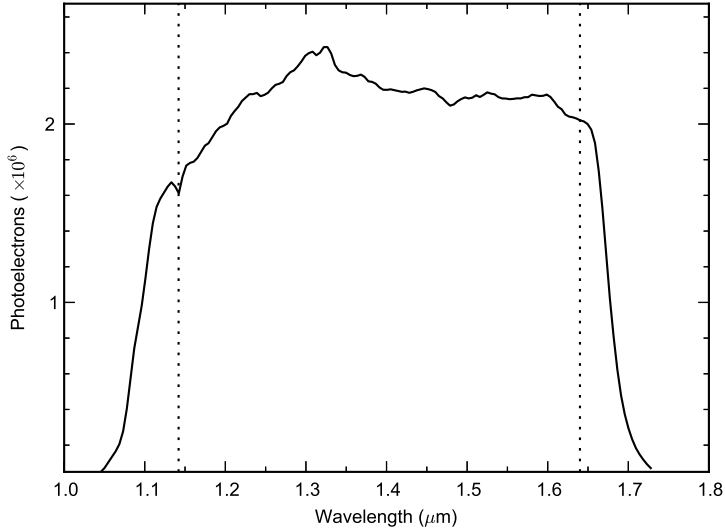}
\caption*{\textbf{Extended Data Figure 2:} An example extracted spectrum for an 88.4\,s exposure.  The dotted lines indicate the wavelength range over
which we measure the transmission spectrum.}
\end{figure*}

Finally, we account for dispersion-direction drift of the spectra during each visit and between visits.  Using the first exposure of the first visit as a template, we determine a shift in wavelength-space that minimizes the difference between each subsequent spectrum and the template.  The best-fit shift values are less than 0.1 pixel, both within each visit and between visits.
We interpolate each spectrum to an average wavelength scale, offset from the template by the mean of the estimated wavelength shifts.  
This step does not have a significant effect on our results.  We bin the spectra in 5-pixel-wide channels, obtaining 29 spectroscopic light curves covering the wavelength range 1.05 -- 1.70\,$\mu$m.  The data near the edges of the grism response curve exhibit more pronounced systematics, so we restrict our analysis to 22 spectroscopic channels between 1.15 and 1.63\,$\mu$m.  The limits are shown in Extended Data Figure 2.

\vspace{-4mm}

\section*{Systematics correction}\vspace{-4mm}
The light curves exhibit a ramp-like systematic similar to that seen in other WFC3 transit spectroscopy data$^{10,18,19}$.  The ramp has a larger amplitude and a different shape in the first orbit compared to subsequent orbits, so we exclude data from the first orbit in our light curve fits, following standard practice.  We correct for systematics in orbits 2 -- 4 using two methods: \\
Method 1: \texttt{model-ramp}\\
This method fits an analytic model to the light curve$^{10}$.  The model has the form:
\begin{equation}
M(\mathbf{t}) = M_{0,\lambda}(\mathbf{t})[C_{\lambda e s} + V_{\lambda e}\mathbf{t_v}][1 - R_{\lambda e o} e^{-\mathbf{t_b}/\tau_\lambda}]
\end{equation}
where $M_{0,\lambda}(\mathbf{t})$ is the model for the systematics-free transit light curve, $\mathbf{t}$ is a vector of observation times, $\mathbf{t_v}$ is a vector with elements $t_{v,i}$ equal to the time elapsed since the first exposure in the visit corresponding to time $t_i$, $\mathbf{t_b}$ is a vector with elements $t_{b,i}$ equal to the time elapsed since the first exposure in the orbit corresponding to time $t_i$, $C_{\lambda e s}$ is a normalization, $V_{\lambda e}$ is a visit-long slope, $R_{\lambda e o}$ is a ramp amplitude, and $\tau_\lambda$ is a ramp timescale. The subscripts $\lambda$, $e$, $s$, and $o$ denote whether a parameter is a function of wavelength, transit epoch, scan direction, and/or orbit number, respectively.\\

Method 2: \texttt{divide-white}\\
The second method assumes the systematics are wavelength-independent and can be modeled with a scaled time series vector of white light curve systematics, denoted $Z(\mathbf{t})$$^{18,20}$. We fit the white light curve $W(\mathbf{t})$ with the \texttt{model-ramp} technique to determine $Z(\mathbf{t})$:
\begin{equation}
Z(\mathbf{t}) = W(\mathbf{t})/M_0(\mathbf{t}),
\end{equation}
where $M_0(\mathbf{t})$ is the best-fit model to the white light curve.  An example white light curve fit, including the $Z$ vector, is shown in Extended Data Figure 3.  

\begin{figure*}[h!]
%\resizebox{\textwidth}{!}
{\includegraphics{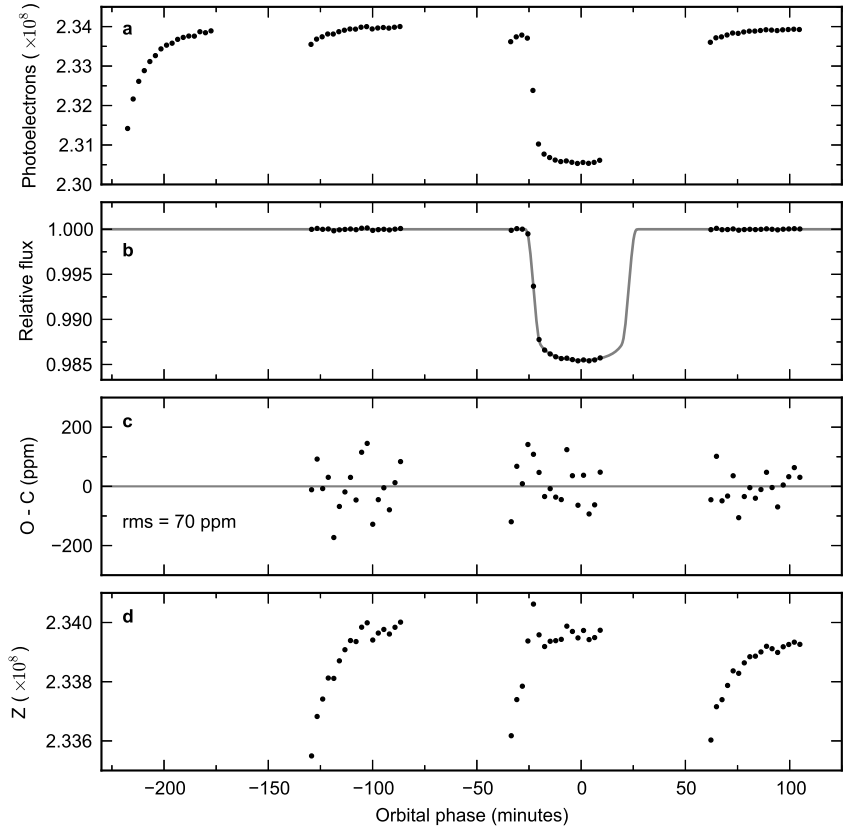}}
\caption*{\textbf{Extended Data Figure 3:} \textbf{a}, The broadband light curve from the first transit observation. \textbf{b}, The broadband light curve corrected for systematics using the \texttt{model-ramp} technique (points) and the best-fit model (line). \textbf{c}, Residuals from the white light curve fit.  \textbf{d}, The vector of systematics $Z$ used in the \texttt{divide-white} technique.}
\end{figure*}

The spectroscopic light curves $S_\lambda(\mathbf{t})$ are modeled as
\begin{equation}
S_\lambda(\mathbf{t}) = C_{\lambda e s}\cdot Z(\mathbf{t}) \cdot S_{0,\lambda}(\mathbf{t})
\end{equation}
where $S_{0,\lambda}(\mathbf{t})$ is the systematics-free transit light curve model for a given wavelength channel and $C_{\lambda e s}$ is a normalization constant.
We observe that the systematics have similar amplitude and form across the wavelength range of our observations, hence the viability of the \texttt{divide-white} technique.
The dominant systematics in our data are related to persistence, which depends on the peak per pixel fluence$^{19}$, but as can be seen in Extended Data Figure 2, the product of the stellar spectrum and the G141 grism response is nearly uniform over the 1.1 -- 1.7\,$\mu$m range.

\vspace{-4mm}

\section*{Light curve fits}\vspace{-4mm}
We fit the spectroscopic light-curves with both the \texttt{divide-white} and \texttt{model-ramp} methods and determined the best-fit parameters and errors with a Markov chain Monte Carlo (MCMC) algorithm.  We divided the light curves into 19 data sets (12 visits, with 2 data sets for 7 of the visits), separated by transit epoch and spatial scan direction, to account for a normalization offset between the forward-scanned and reverse-scanned light curves.  We fit the data sets in each spectral channel with five $10^5$ step MCMC chains, with $2.5\times10^4$ burn-in steps removed from each chain.  We tested for convergence using the Gelman-Rubin diagnostic.  The results reported are from the five chains combined.  

We analyzed each spectral channel independently.  The free parameters for the \texttt{divide-white} fit are a normalization constant $C_{\lambda e s}$, a linear limb darkening parameter $u_\lambda$, and the planet-to-star radius ratio $R_p/R_{s,\lambda, e}$.  The \texttt{model-ramp} fit had these same free parameters, plus an additional visit-long slope parameter $V_{\lambda e}$, ramp amplitudes $R_{\lambda e o}$, and a ramp timescale $\tau_\lambda$. We constrained the ramp amplitudes for orbits 3 and 4 to be equal within each data set.  For both methods, we held the following orbital parameters fixed at the best-fit values for the white light curve: inclination $i = 89.1^\circ$, the ratio of the semi-major axis to the stellar radius $a/R_s = 15.23$, the orbital period $P =1.58040464894$\,days, and the time of central transit $T_c = 2454966.52488$ $\mathrm{BJD_{TDB}}$.  We assume a circular orbit.  There were a total of 32 and 67 free parameters per channel for the \texttt{divide-white} and \texttt{model-ramp} fits, respectively.
The priors for each free parameter were uniform.  We checked that the light curves are sufficiently precise to fit for all the free parameters by visually inspecting pairs plots for the fit parameters.  As an example, we show the posterior distributions of the parameters for the \texttt{divide-white} fit of the 1.40 $\mu$m channel in Extended Data Figure 4, and note that there is little correlation between parameters.

We report the measured transit depths, limb darkening parameters, and $\chi^2_\nu$ values for both the \texttt{divide-white} and \texttt{model-ramp} methods in Extended Data Table 1.  The transit depths given are the weighted averages over all epochs, minus the mean transit depth over all channels (0.013490 for \texttt{divide-white} and 0.013489 for \texttt{model-ramp}).  
For the results given in the main text, we use the \texttt{divide-white} spectrum because the light curve fits from this method have fewer free parameters and lower $\chi^2_\nu$ values.  

We performed several consistency checks as part of our analysis, enumerated below.
\begin{enumerate}
\item{We verified that the transmission spectra obtained with the \texttt{divide-white} and \texttt{model-ramp} methods are consistent within 1\,$\sigma$, and that the main conclusions of the paper are not affected by which method we chose.}
\item{We confirmed that the measured transit depths are consistent from epoch to epoch, as shown in Extended Data Figure 5.  As a test, we fit separate transit depths to the forward- and reverse-scanned data, and found that the transit depths are consistent for the two scan directions.}
\item{We tested the effects of using an inaccurate white light curve model ($M_0$) for the \texttt{divide-}\texttt{white} method and found that the results are robust to changes in $M_0$.  For example, changing the model transit depth by 5\,$\sigma$ from the best-fit white light curve value affects the relative spectroscopic transit depths by less than 1 ppm.} 
\item{We compare our results to the previously published WFC3 transmission spectrum for GJ\,1214b$^{10}$.  Our relative transit depths are within 1\,$\sigma$ for 18 wavelength channels and within 2\,$\sigma$ for the other four channels.  The 2\,$\sigma$ differences are not clustered in wavelength.}
\end{enumerate}

We also considered the effects of stellar activity on the transmission spectrum and found that it does not impact our results.
The measured transit depths are consistent over all epochs in the white light curve and the spectroscopic light curves, which suggests that the influence of stellar activity on the spectrum is minimal.  To confirm this, we simulated the effect of star spots\cite{desert11} assuming they are 300\,K cooler than the 3250\,K stellar photosphere$^{22}$, and find that their influence is below our measurement precision.
We also considered the possibility that the star spots have excess water due to their cooler temperature. This could introduce a water feature in the transmission spectrum, but it would not cancel out water features from the planet's atmosphere.  Given that we do not see evidence for water absorption in the spectrum, any contribution from water in unocculted star spots must be below the level of precision in our data.  As a check, we computed the transmission spectrum from three transits occurring over a timespan of just two weeks, during which time the spot coverage should be roughly constant.  Even with just these three transits, we rule out a pure H$_2$O atmosphere at $>5$\,$\sigma$ confidence, which confirms that any noise introduced by unocculted spots does not change our conclusion.

The derived limb darkening coeffients are shown in Extended Data Figure 6.  We fit a single limb darkening coefficient to each channel and constrained the value to be the same for all the transits.  
Our limb darkening fits illustrate the importance of careful treatment of limb darkening for cool stars.
There is a peak in the coefficients near $1.45\,\mu$m that
is due to the presence of water in the star.
As a result of this, fixing the limb darkening coefficients to a constant value in all spectral channels introduces a spurious water feature in the transmission spectrum.  However, we can fit the limb darkening precisely with our data and it is non-degenerate with the transit depth.  The uncertainty in our limb darkening fits introduces an uncertainty in the transit depth of less then 1 ppm.
To confirm that fitting a linear limb darkening parameter is appropriate for our data, we simulated a data set using a model water vapor atmosphere, quadratic limb darkening coefficents from a 3250 K stellar model, and the residuals and systematics from the real data.  We analyze this mock data set in the same way we treat the real data and find that we fully recover the water vapor transmission spectrum with a single limb darkening parameter.

%\singlespacing

\begin{table}
\centering
\caption{\textbf{Derived parameters for the light curve fits for the \texttt{divide-white} (\texttt{d-w}) and \texttt{model-ramp} (\texttt{m-r}) techniques}}
	\begin{tabular}[c]{cr@{}c@{}lr@{}c@{}lcccc}
	\hline \\ [-1.5ex]
	Wavelength ($\mu$m) & \multicolumn{6}{c}{Transit Depth (ppm)} & \multicolumn{2}{c}{Limb Darkening} &\multicolumn{2}{c}{$\chi^2_\nu$} \\
	\, & \multicolumn{3}{c}{\texttt{d-w}} & \multicolumn{3}{c}{\texttt{m-r}} & \texttt{d-w} & \texttt{m-r} & \texttt{d-w} & \texttt{m-r} \\
	%\, & \, & \, & \, & \, & \,& \,&\,&\,&622 dof & 597 dof \\
	\hline \\ [-1.5ex]
	$1.135 - 1.158$ & $-39\,$ & $\pm$ & $\,31$ & $6\,$ & $\pm$ & $\,33$ & $0.27 \pm 0.01$ & $0.28 \pm 0.01$ & $1.12$ & $1.20$ \\
	$1.158 - 1.181$ & $-28\,$ & $\pm$ & $\,30$ & $12\,$ & $\pm$ & $\,32$ & $0.26 \pm 0.01$ & $0.27 \pm 0.01$ & $1.01$ & $1.24$ \\
	$1.181 - 1.204$ & $34\,$ & $\pm$ & $\,30$ & $29\,$ & $\pm$ & $\,30$ & $0.25 \pm 0.01$ & $0.26 \pm 0.01$ & $1.04$ & $1.44$ \\
	$1.205 - 1.228$ & $-48\,$ & $\pm$ & $\,28$ & $-32\,$ & $\pm$ & $\,29$ & $0.26 \pm 0.01$ & $0.28 \pm 0.01$ & $0.90$ & $1.22$ \\
	$1.228 - 1.251$ & $27\,$ & $\pm$ & $\,28$ & $25\,$ & $\pm$ & $\,29$ & $0.26 \pm 0.01$ & $0.28 \pm 0.01$ & $0.85$ & $1.29$ \\
	$1.251 - 1.274$ & $5\,$ & $\pm$ & $\,27$ & $-6\,$ & $\pm$ & $\,29$ & $0.26 \pm 0.01$ & $0.26 \pm 0.01$ & $0.97$ & $1.29$ \\
	$1.274 - 1.297$ & $13\,$ & $\pm$ & $\,27$ & $12\,$ & $\pm$ & $\,27$ & $0.23 \pm 0.01$ & $0.23 \pm 0.01$ & $1.00$ & $1.50$ \\
	$1.297 - 1.320$ & $14\,$ & $\pm$ & $\,26$ & $0\,$ & $\pm$ & $\,27$ & $0.23 \pm 0.01$ & $0.25 \pm 0.01$ & $0.96$ & $1.38$ \\
	$1.320 - 1.343$ & $29\,$ & $\pm$ & $\,26$ & $2\,$ & $\pm$ & $\,28$ & $0.26 \pm 0.01$ & $0.27 \pm 0.01$ & $1.08$ & $1.52$ \\
	$1.343 - 1.366$ & $-2\,$ & $\pm$ & $\,27$ & $-15\,$ & $\pm$ & $\,28$ & $0.30 \pm 0.01$ & $0.32 \pm 0.01$ & $0.99$ & $1.44$ \\
	$1.366 - 1.389$ & $32\,$ & $\pm$ & $\,27$ & $35\,$ & $\pm$ & $\,26$ & $0.28 \pm 0.01$ & $0.29 \pm 0.01$ & $0.97$ & $1.42$ \\
	$1.389 - 1.412$ & $31\,$ & $\pm$ & $\,27$ & $33\,$ & $\pm$ & $\,28$ & $0.28 \pm 0.01$ & $0.29 \pm 0.01$ & $0.96$ & $1.39$ \\
	$1.412 - 1.435$ & $-5\,$ & $\pm$ & $\,27$ & $-33\,$ & $\pm$ & $\,28$ & $0.29 \pm 0.01$ & $0.31 \pm 0.01$ & $1.15$ & $1.51$ \\
	$1.435 - 1.458$ & $29\,$ & $\pm$ & $\,29$ & $17\,$ & $\pm$ & $\,28$ & $0.29 \pm 0.01$ & $0.30 \pm 0.01$ & $1.01$ & $1.39$ \\
	$1.458 - 1.481$ & $-8\,$ & $\pm$ & $\,28$ & $1\,$ & $\pm$ & $\,29$ & $0.32 \pm 0.01$ & $0.33 \pm 0.01$ & $1.01$ & $1.33$ \\
	$1.481 - 1.504$ & $27\,$ & $\pm$ & $\,28$ & $28\,$ & $\pm$ & $\,28$ & $0.28 \pm 0.01$ & $0.29 \pm 0.01$ & $0.94$ & $1.37$ \\
	$1.504 - 1.527$ & $-11\,$ & $\pm$ & $\,28$ & $-23\,$ & $\pm$ & $\,29$ & $0.27 \pm 0.01$ & $0.29 \pm 0.01$ & $1.15$ & $1.58$ \\
	$1.527 - 1.550$ & $20\,$ & $\pm$ & $\,28$ & $1\,$ & $\pm$ & $\,29$ & $0.27 \pm 0.01$ & $0.29 \pm 0.01$ & $1.17$ & $1.56$ \\
	$1.550 - 1.573$ & $-21\,$ & $\pm$ & $\,28$ & $0\,$ & $\pm$ & $\,28$ & $0.28 \pm 0.01$ & $0.29 \pm 0.01$ & $1.20$ & $1.62$ \\
	$1.573 - 1.596$ & $-65\,$ & $\pm$ & $\,28$ & $-62\,$ & $\pm$ & $\,30$ & $0.26 \pm 0.01$ & $0.28 \pm 0.01$ & $1.08$ & $1.46$ \\
	$1.596 - 1.619$ & $-17\,$ & $\pm$ & $\,28$ & $-6\,$ & $\pm$ & $\,29$ & $0.26 \pm 0.01$ & $0.27 \pm 0.01$ & $1.34$ & $1.69$ \\
	$1.619 - 1.642$ & $-17\,$ & $\pm$ & $\,30$ & $-26\,$ & $\pm$ & $\,30$ & $0.22 \pm 0.01$ & $0.24 \pm 0.01$ & $1.16$ & $1.59$ \\
	\hline
	\end{tabular}
\end{table}

\begin{figure*}[h!]
{\includegraphics{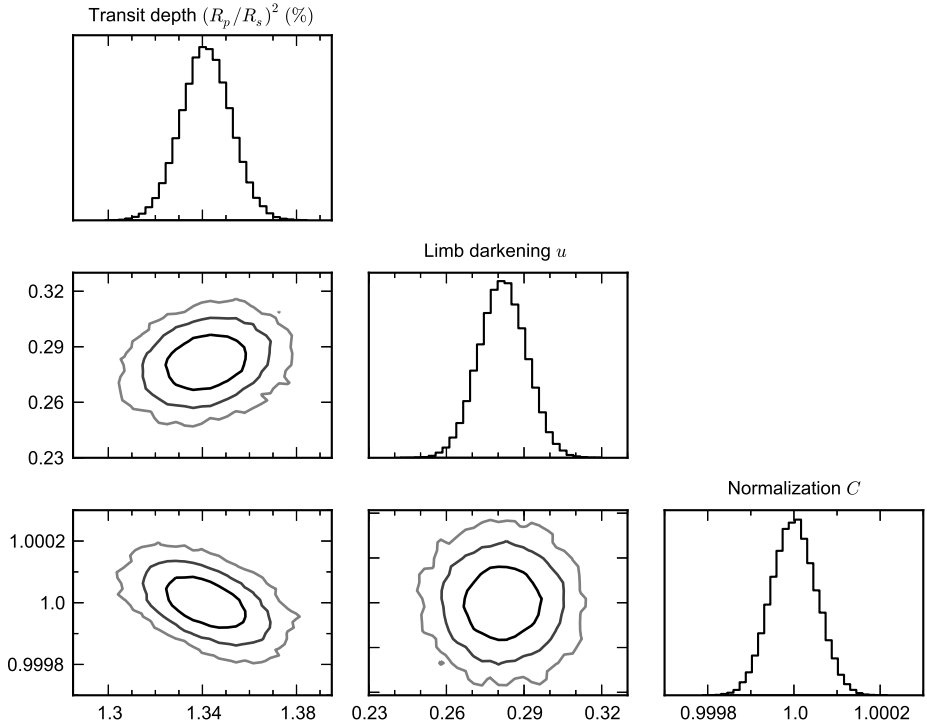}}
\caption*{\textbf{Extended Data Figure 4:} The posterior distributions for the \texttt{divide-white} fit parameters for the 1.40\,$\mu$m channel from the first transit observation.  The diagonal panels show histograms of the Markov chains for each parameter.  The off-diagonal panels show contour plots for pairs of parameters, with lines indicating the 1, 2, and 3\,$\sigma$ confidence intervals for the distribution.  The normalization constant is divided by its mean.}
\end{figure*}

\begin{figure*}[h!]
{\includegraphics{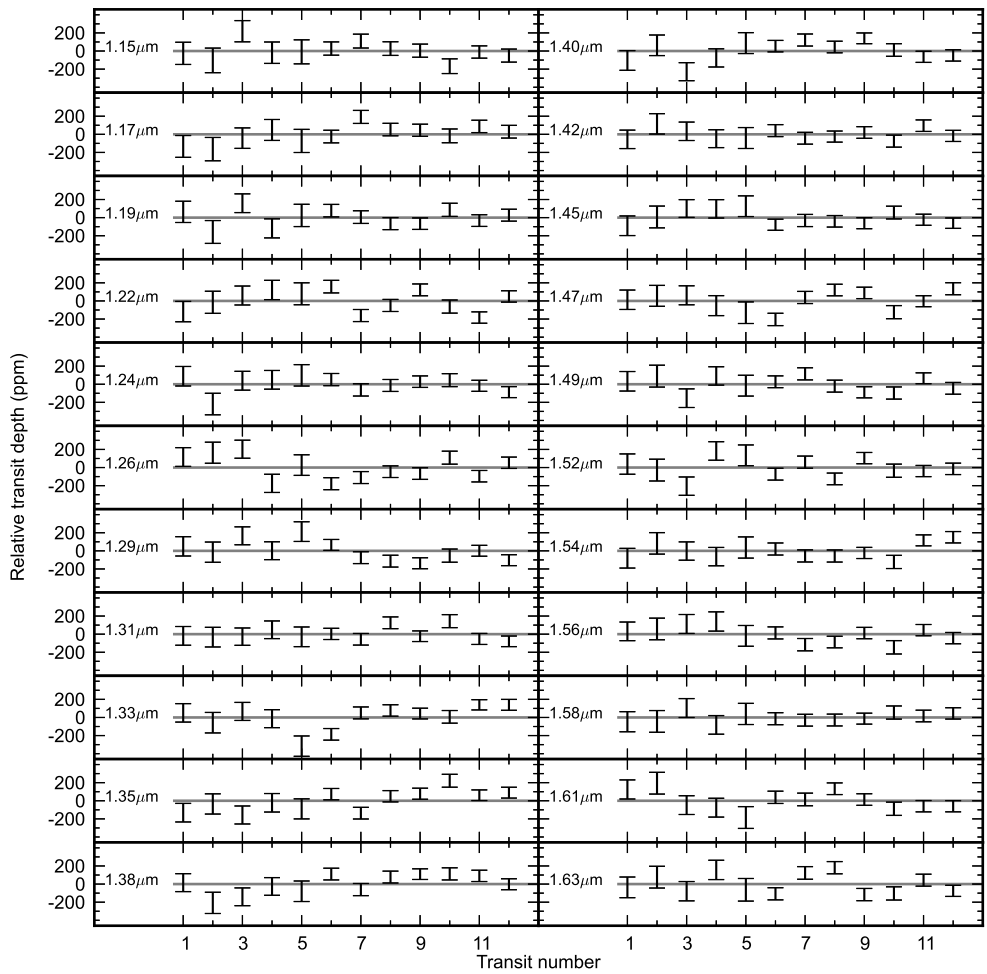}}
\caption*{\textbf{Extended Data Figure 5:} Transit depths relative to the mean in 22 spectroscopic channels, for the 12 transits analyzed.
The black error bars indicate the 1\,$\sigma$ uncertainties determined by MCMC.}
\end{figure*}

\begin{figure*}[h!]
{\includegraphics{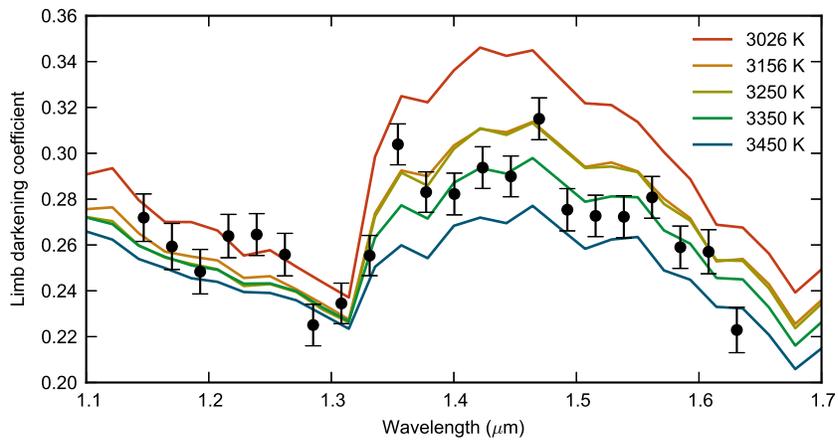}}
\caption*{\textbf{Extended Data Figure 6:} Fitted limb darkening coefficients as a function of wavelength (black points) and
theoretical predictions for stellar atmospheres with a range of temperatures (lines).  The uncertainties are
1\,$\sigma$ confidence intervals from an MCMC.  The temperature of GJ\,1214 is estimated to be 3250 K\cite{anglada13}.}
\end{figure*}

\end{document}